\documentclass[iop]{emulateapj}

\usepackage{natbib}  
\usepackage{color}
\usepackage{epsfig,graphicx,graphics}

\shorttitle{Intra-cavity kinematics}
\shortauthors{Casassus et al.}

\begin{document}


\title{Accretion kinematics through the warped transition disk in HD~142527
from resolved CO(6-5) observations}


\author{S. Casassus\altaffilmark{1,2}, S. Marino\altaffilmark{1,2},
  S. P\'{e}rez\altaffilmark{1,2}, P. Roman\altaffilmark{3,2},
  A. Dunhill\altaffilmark{4,2}, P.J. Armitage\altaffilmark{5}, J. Cuadra\altaffilmark{4,2}, 
  A. Wootten\altaffilmark{6}, G. van der Plas\altaffilmark{1,2},
  L. Cieza\altaffilmark{7,2}, Victor Moral\altaffilmark{3,2},
  V. Christiaens\altaffilmark{1,2}, Mat\'{\i}as
  Montesinos\altaffilmark{1,2}}

%
%
%


\altaffiltext{1}{Departamento de Astronom\'{\i}a, Universidad de Chile, Casilla 36-D, Santiago, Chile}
\altaffiltext{2}{Millennium Nucleus ``Protoplanetary Disks'', Chile}
\altaffiltext{3}{Center for Mathematical Modeling, Universidad de Chile,  Av. Blanco Encalada 2120 Piso 7, Santiago, Chile}
\altaffiltext{4}{Instituto de Astrof\'{\i}sica, Pontificia Universidad Cat\'olica de Chile, 7820436 Macul, Santiago,
  Chile}
\altaffiltext{5}{JILA, University of Colorado and NIST, UCB 440, Boulder, CO 80309, USA}
\altaffiltext{6}{National Radio Astronomy Observatory, 520 Edgemont Road, Charlottesville, VA 22903-2475, USA}
\altaffiltext{7}{Facultad de Ingenier\'{\i}a,  Universidad Diego Portales, Av. Ej\'ercito 441, Santiago, Chile}


\begin{abstract}
The finding of residual gas in the large central cavity of the
HD~142527 disk motivates questions on the origin of its non-Keplerian
kinematics, and possible connections with planet formation.  We aim to
understand the physical structure that underlies the intra-cavity
gaseous flows, guided by new molecular-line data in CO(6-5) with
unprecedented angular resolutions. Given the warped structure inferred
from the identification of scattered-light shadows cast on the outer
disk, the kinematics are consistent, to first order, with axisymmetric
accretion onto the inner disk occurring at all azimuth.  A
steady-state accretion profile, fixed at the stellar accretion rate,
explains the depth of the cavity as traced in CO isotopologues. The
abrupt warp and evidence for near free-fall radial flows in HD 142527
resemble theoretical models for disk tearing, which could be driven by
the reported low mass companion, whose orbit may be contained in the
plane of the inner disk. The companion's high inclination with respect
to the massive outer disk could drive Kozai oscillations over long
time-scales; high-eccentricity periods may perhaps account for the
large cavity. While shadowing by the tilted disk could imprint an
azimuthal modulation in the molecular-line maps, further observations
are required to ascertain the significance of azimuthal structure in
the density field inside the cavity of HD~142527.
\end{abstract}





\section{Introduction}\label{sec:intro}

While in its first year of Early Science operations, the Atacama Large
Millimeter Array (ALMA) already provided resolved sub-mm observations
of circumstellar disks around young stars, where planet formation must
occur.  The pre-main-sequence star HD~142527 (F6{\sc iii}e, at a
distance of $\sim$145~pc) is particularly interesting as its disk is viewed
close to face-on, and its protoplanetary cavity is amenable to
scrutiny thanks to its record size
\citep{2006ApJ...636L.153F}. Indeed, ALMA Cycle~0 observations of
HD~142527 \citep[][]{Casassus2013Natur, Perez2015ApJ...798...85P}
revealed gaseous flows inside the dust-depleted central cavity, which
is an expected feature of the formation of planetary systems
\citep{Zhu2011,2011ApJ...738..131D}. Here we complete this ALMA
program with high-frequency molecular-line data that provide
additional insight on the physical structures giving rise to the
intra-cavity flows in HD~142527.

An obvious feature of the first ALMA observations of HD~142527 is the
strong and centrally peaked HCO$^+$(4-3) emission, which is consistent
with radial flows in addition to Keplerian rotation
\citep{Casassus2013Natur}. In this scenario the mass transfer rate
inferred from the HCO$^+$ kinematics matches the stellar accretion
rate. It is tempting to associate the HCO$^+$ kinematics with
planet-induced accretion flows, i.e. gaseous streamers due to as yet
undetected protoplanets inside the cavity. Such accretion kinematics
should have counterparts in other molecular tracers, and indeed the
velocity centroid map in CO(3-2) also hints at non-Keplerian flows
\citep{Casassus2013Natur, Casassus2013rocks,
  Rosenfeld2014ApJ...782...62R}.  However, the coarse Cycle~0 beam in
CO(3-2), along with absorption by a diffuse CO screen along the line
of sight to HD~142527, prevent adequate sampling of intra-cavity
kinematics.  Additionally a warped disk instead of a stellocentric
velocity component could result in the same CO(3-2) centroid maps
\citep{Rosenfeld2014ApJ...782...62R}.

The new ALMA observations we present here, acquired in the CO(6-5)
line, together with a well-determined inner disk orientation thanks to
the identification of its shadows in
scattered-light \citep{Marino2015ApJ...798L..44M}, allow us to conclude
that the intra-cavity flows do correspond to stellocentric accretion
occurring at all azimuth, and near free-fall
velocities. Section~\ref{sec:obs} presents the observational results
on the intra-cavity gas kinematics. Sec.~\ref{sec:model} introduces
the parametric models we use to interpret the data, and compares the
models with the observations.  Sec.~\ref{sec:discussion} discusses
implications of the inferred kinematics in the light of the recent
theoretical discovery of ``disk tearing'' in strongly warped
disks \citep{Nixon_2013MNRAS.434.1946N}.  Sec.~\ref{sec:conclusion}
summarizes our conclusions on the kinematics of stellocentric
accretion in the cavity of HD~142527.

\section{Observations} \label{sec:obs}

HD~142527 was observed with ALMA in band~9, at $\sim$700~GHz, during
June 2012 and with baselines ranging from 0.04~M$\lambda$ to
0.9~M$\lambda$.  We refer to Appendix~\ref{sec:data} for an account of
data reduction.

\subsection{Choice of image-plane representations of the visibility data}

Image synthesis was performed with a non-parametric least-squares fit
to the visibility data (see Appendix~\ref{sec:imagesynthesis}); the
resulting model images are labelled `MEM'. These `deconvolved' images
pick up finer details than the conventional `restored' maps, which are
produced by convolving the MEM maps with an ellipical Gaussian called
`clean' beam, which is fit to the `dirty' beam (or point-spread-function,
PSF). With the addition of a dirty map of the model residuals, the
restored maps allow a standard representation of the data, with a
single and well defined angular resolution.  However, the moments
extracted on the `restored' maps (Appendix~\ref{sec:imagesynthesis})
are affected by convolution with the clean beam, which enhances
velocity dispersions by aligning different velocity components, with
consequences on the intensity maps if extracting moments with Gaussian
spectral fits (see Appendix~\ref{sec:synthesis_compar}). We thus chose
to base our study on the moments maps inferred from deconvolved (MEM)
datacubes. A drawback of this choice is that since we are
extrapolating power in the Fourier domain, the resolution of the MEM
images is variable and depends on the image. Here the full-width at
half-maximum (FWHM) of the MEM PSF varies from 0.062\arcsec~ $\times$
0.051\arcsec~ for a bright point source, up to $\sim$50\% coarser for
fainter extended emission (approximated by a collection of spikes).

\subsection{CO(6-5) kinematics}

\begin{figure*}
\begin{center}
  \includegraphics[width=\textwidth,height=!]{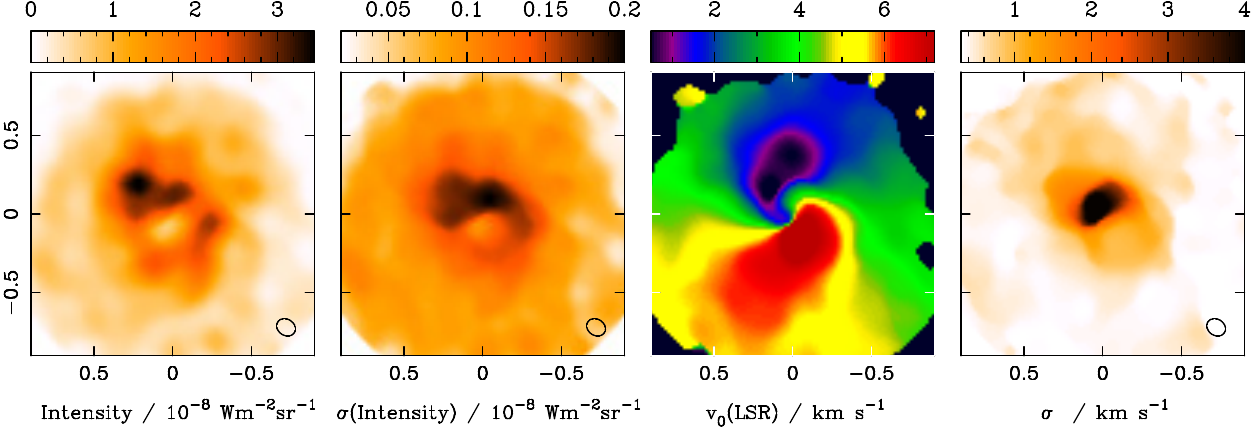}
\end{center}
\caption{Moment maps in CO(6-5) emission. $x-$ and $y-$axis show
  angular offset along RA and DEC from the stellar position, in
  arcsec. The moments were extracted on deconvolved MEM models. From
  left to right we show frequency-integrated line intensity,
  uncertainty on the intensity field, velocity centroid, and
  1~$\sigma$ velocity dispersion.  \label{fig:co65moments_memonly}}
\end{figure*}

The main properties of the continuum-subtracted CO(6-5) data are
summarised by the moment maps shown in
Fig.~\ref{fig:co65moments_memonly}.  The CO(6-5) line intensity
displays a central decrement so that the minimum intensity is a factor
of $\sim 0.25\pm0.03$ times lower than the average intensity in the
brighter ring at a radius of 0.2\arcsec. This central decrement is
probably related to photodissociation of CO, so that the very central
CO(6-5) could be rarefied and broad enough to be optically thin. The
more extended emission in CO(6-5) seems to be ascribed to the
dust-depleted cavity, with a sharp drop at $\sim$1~arcsec,
corresponding to the radius of the Bright crescent-shaped outer ring
seen in the continuum.  This sharp drop is absent in CO(3-2), which
extends well outside the cavity, so that the CO temperature outside
the cavity is probably not high enough to excite $J=6$.

The disk position angle (PA) is defined as the intersection of the
plane of a disk with the plane of the sky. For Keplerian rotation in
disks with axial symmetry, the velocity centroid maps bear reflection
symmetry about the disk PA, and the highest Doppler-shifted emission
towards the blue and red are aligned along the PA. In
Fig.~\ref{fig:co65moments_memonly} the velocity centroid is more
complex than the usual butterfly pattern. It is as if the velocity
field had been `twisted' anti-clockwise, as would happen if the disk
PA rotated with stellocentric radius $r$, from $-20$\degr\, East of
North outside $ r \sim 0.6$\arcsec, to $\sim +60$\degr\, at
$r\sim$0.25\arcsec, coincident with the extremes in velocity
centroid. Inside 0.1\arcsec, the PA appears to rotate further and
reaches $110\pm10$\degr\,.  Yet the scattered light shadows from the
inner disk show that its PA is $-8\pm5$\degr\,
\citep[][]{Marino2015ApJ...798L..44M}.



The CO(6-5) velocity dispersion reaches very high values, especially
near the star. Velocity dispersion values above 4~km~s$^{-1}$ are
found in the unresolved central beam; the bulk of this dispersion is
probably due to flow kinematics. However, values of about
2~km~s$^{-1}$ are also found away from the central beam, which persist
after deconvolution. For instance, in Fig.~\ref{fig:co65moments_memonly} the
1~$\sigma$ velocity dispersion in the restored maps, $\sigma_{\rm
  rest.}$, measured at an offset $\vec{P}= (+0.2\arcsec,+0.2\arcsec)$,
is 2.01~km~s$^{-1}$, while the 1~$\sigma$ dispersion extracted from
the deconvolved datacubes, $\sigma_{\rm mem}$, still shows a
dispersion of 1.78~km~s$^{-1}$. Further tests on the suitability of
the deconvolved images as measures of the intrinsic line-width are
provided in Appendix~\ref{sec:disp}.

\begin{figure*}
\begin{center}
\includegraphics[width=0.9\textwidth,height=!]{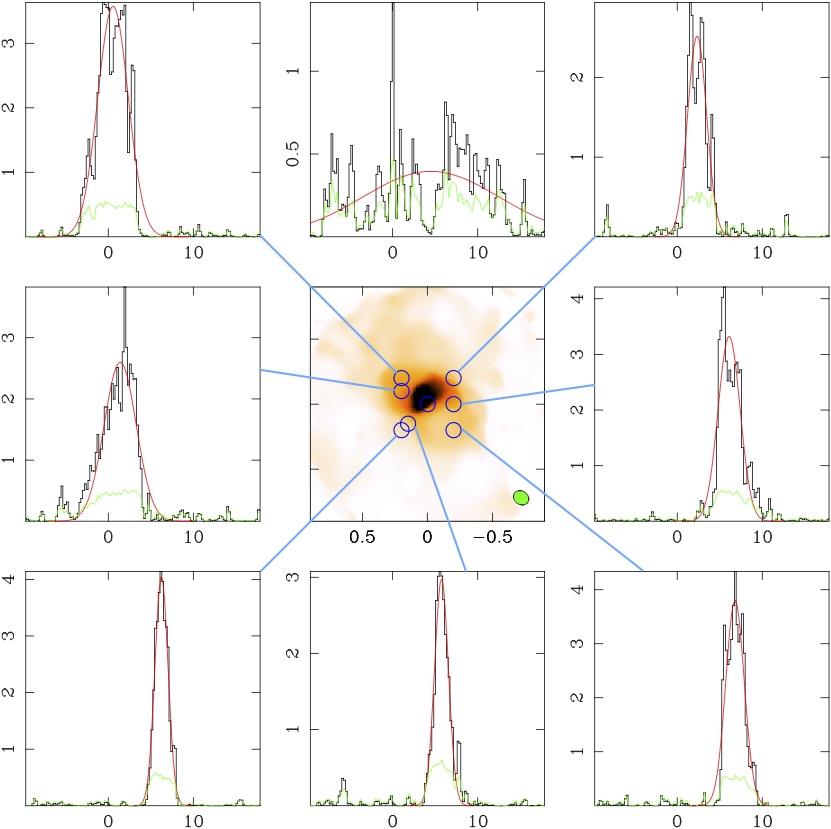}
\end{center}
\caption{Example CO(6-5) line profiles. We select 8 locations near the
  star to illustrate the line profiles obtained in the deconvolved
  datacube. The central plot shows $\sigma_{\rm mem}$, and is repeated
  from Fig.~\ref{fig:co65moments_memonly}. The spectra are extracted
  from single pixels in the deconvolved datacube. Solid black is used
  to plot specific intensity in $10^{11}$~Jy~sr$^{-1}$, along the
  $y-$axis, as a function of LSR velocity in $x-$axis, in
  km~s$^{-1}$. Solid green plots the 1~$\sigma$ error on the specific
  intensity spectra, as estimated from Monte-Carlo runs (see
  Sec.~\ref{sec:imagesynthesis}).  \label{fig:los_spec}}
\end{figure*}

Beyond 0.1\arcsec~ from the nominal stellar position, the line
profiles appear to be singly peaked, as illustrated in
Fig.~\ref{fig:los_spec}. If due to thermal broadening, the CO velocity
distribution corresponds to temperatures $T = 28 m_p \sigma^2 /k$ in
excess of 5000~K, as far out as 0.5\arcsec~ from the star.  Another
mechanism for producing the observed line-widths is microscopic
turbulence.  There are however hints of line asymmetry, such as
shoulders or lack of low-level Gaussian wings, which suggest the
presence of multiple velocity flows stratified along the line of
sight. However the significance of these spectral features is not high
enough to motivate detailed interpretation.

%

\begin{figure}
\begin{center}
\includegraphics[width=0.8\columnwidth,height=!]{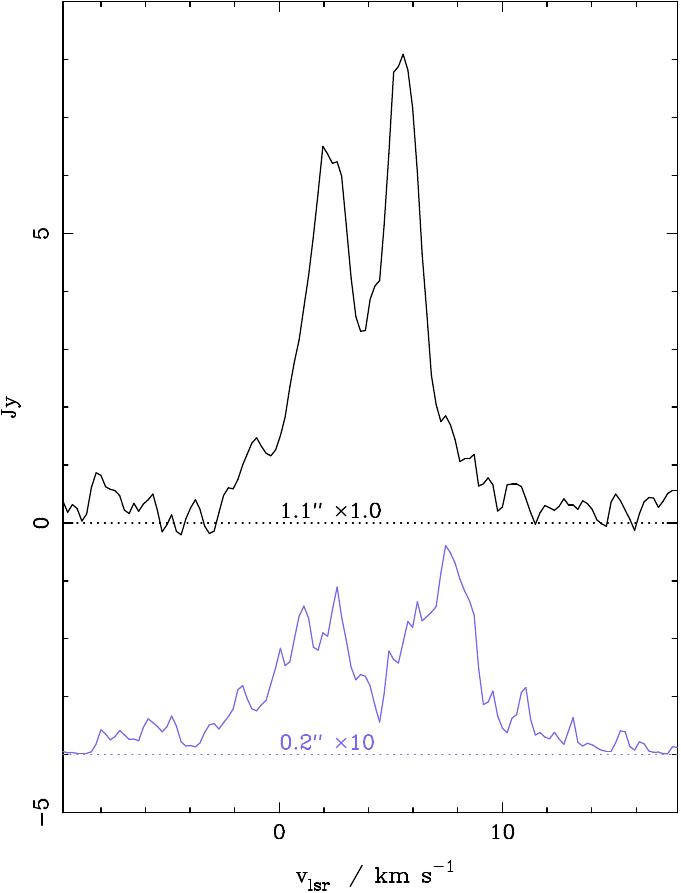}\\
\includegraphics[width=0.8\columnwidth,height=!]{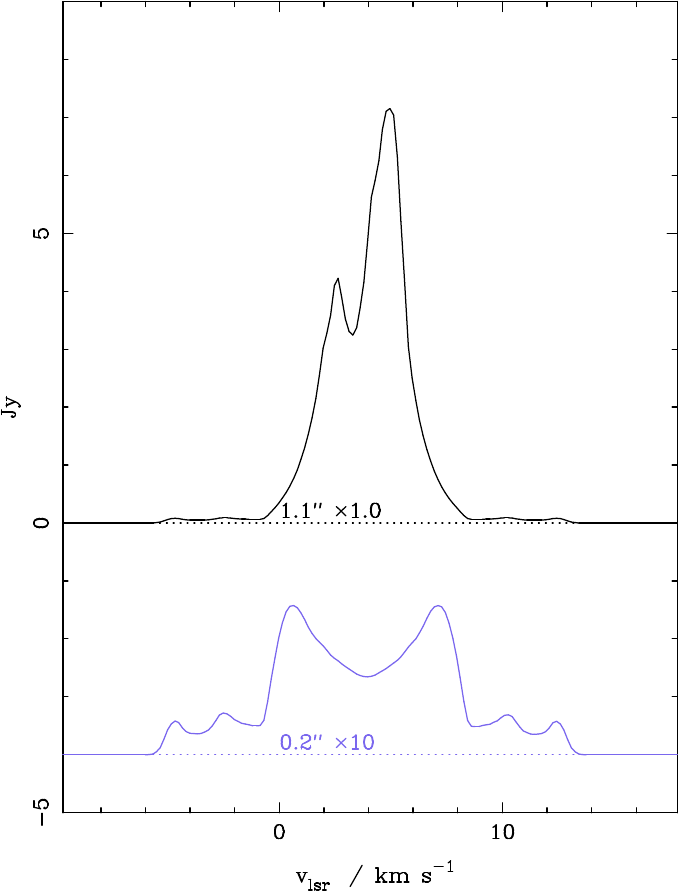}
\end{center}
\caption{Line profiles in CO(6-5). {\bf Top:} Observed profile,
  extracted on the restored datacube in natural weights, using
  aperture photometry inside a 1.1\arcsec~ radius (solid black), and
  also in the {\tt uvmem} model with a 0.2\arcsec~ radius (in solid
  blue, magnified by a factor of 10 and offset for clarity). The
  horizontal dotted lines indicate the origins of flux density for
  each extraction.  {\bf Bottom:} Corresponding line profiles
  extracted on the radiative transfer predictions described in
  Sec.~\ref{sec:model}. The wiggles in the line wings have
  counterparts in the observations - this may be better appreciated by
  inspection of the channel maps in Appendix~\ref{sec:data}.  
  \label{fig:spectra} }
\end{figure}

Integrated line profiles for CO(6-5) are shown in
Fig.~\ref{fig:spectra}. The total line flux extracted from the
restored datacube, using natural weights, and in a circular aperture
with a radius of 2.5\arcsec, is 70~Jy~km~s$^{-1}$. This value is
fairly close to the flux density of 80~Jy~km~s$^{-1}$ reported by
\citet{Casassus2013A&A...553A..64C} from APEX+CHAMP+ single-dish
observations.  The double-peaked CO(6-5) profiles obtained with ALMA
are typical of Keplerian disks, and are also reminiscent of the CHAMP+
spectrum, given its noise level. However, here the blue-shifted peak
appears to be fainter than the red-shifted peak by $\sim$20\%. This
line asymmetry is opposite to that seen in $^{12}$CO(2-1)
\citep{Perez2015ApJ...798...85P}, but is similar to HCO$^+$(4-3)
\citep{Casassus2013Natur} and the CO isotopologues
\citep{Perez2015ApJ...798...85P}. It is likely that the lower-J
$^{12}$CO is absorbed by the diffuse screen at red-shifted velocities
\citep{Casassus2013A&A...553A..64C}. The origin of the blue-shifted
decrement in species not affected by the diffuse molecular screen is
probably related to the anti-correlation with the continuum: in
continuum-subtracted maps, the line emission appears to be quenched
under the crescent-shaped continuum \citep{Casassus2013Natur,
  vdP_2014ApJ...792L..25V, Perez2015ApJ...798...85P}, at locations
that correspond to blue-shifted velocities.  The pointing offset
invoked by \citet{Casassus2013A&A...553A..64C} to account for the
asymmetric CHAMP+ profile is probably unnecessary given the match in
line flux with the interferometer CO(6-5) maps, and given the later
finding of the molecular decrements at blue-shifted velocities.

\subsection{Complementary sub-mm molecular-line data} \label{sec:b7}

As mentioned in Sec.~\ref{sec:intro} the HCO$^+$(4-3) and CO(3-2) data
previously described in the literature also display deviations from
Keplerian rotation with a constant disk orientation.  Although the
outer disk follows Keplerian rotation with a PA of $-20$\degr, inside
the cavity the PA joining regions of opposite
velocities\footnote{relative to systemic} rotates with stellocentric
distance resulting in `twisted' kinematics \citep{Casassus2013rocks}.
The high velocity\footnote{with velocities corresponding to $
  v_\mathrm{lsr} / $~km~s$^{-1} \in [-2.8,3.2]$ and $\in [4.8, 8.8]$,
  while the systemic velocity is at $v_\mathrm{lsr}
  =+3.8\pm0.1~$km~s$^{-1}$} blue and red-shifted HCO$^+$ are offset by
0.2~arcsec, with a PA of $\sim$60\degr\, \citep{Casassus2013Natur}, while
the outer disk PA is at $\sim-$20\degr.  Low-velocity filamentary
HCO$^+$(4-3), within 0.8~km~s$^{-1}$ of systemic, connects the outer
disk to the East with the central broad-velocity HCO$^+$
\citep{Casassus2013Natur}, along a PA that coincides with that of the
blue-shifted signal.  A similar but fainter filament may connect with
the Western outer disk.  Interestingly the peak CO(3-2) intensity is
offset from the star, again towards the same direction as the
blue-shifted HCO$^+$. The PA joining the blue- and red-shifted
velocities is thus $\sim$60\degr\, and common for CO(3-2) and HCO$^+$, and
approximately orthogonal to the PA of the outer disk.

Such kinematic features, seen as extremes in HCO$^+$, and probably
also in CO(3-2), are expected in the framework of an additional radial
velocity component directed towards the star. In order to impact the
observed kinematics, the additional radial velocity component must be
comparable in magnitude to the rotation velocity
\citep{Casassus2013rocks}, so close to free-fall.

\begin{figure*}
\begin{center}
\includegraphics[width=0.8\textwidth,height=!]{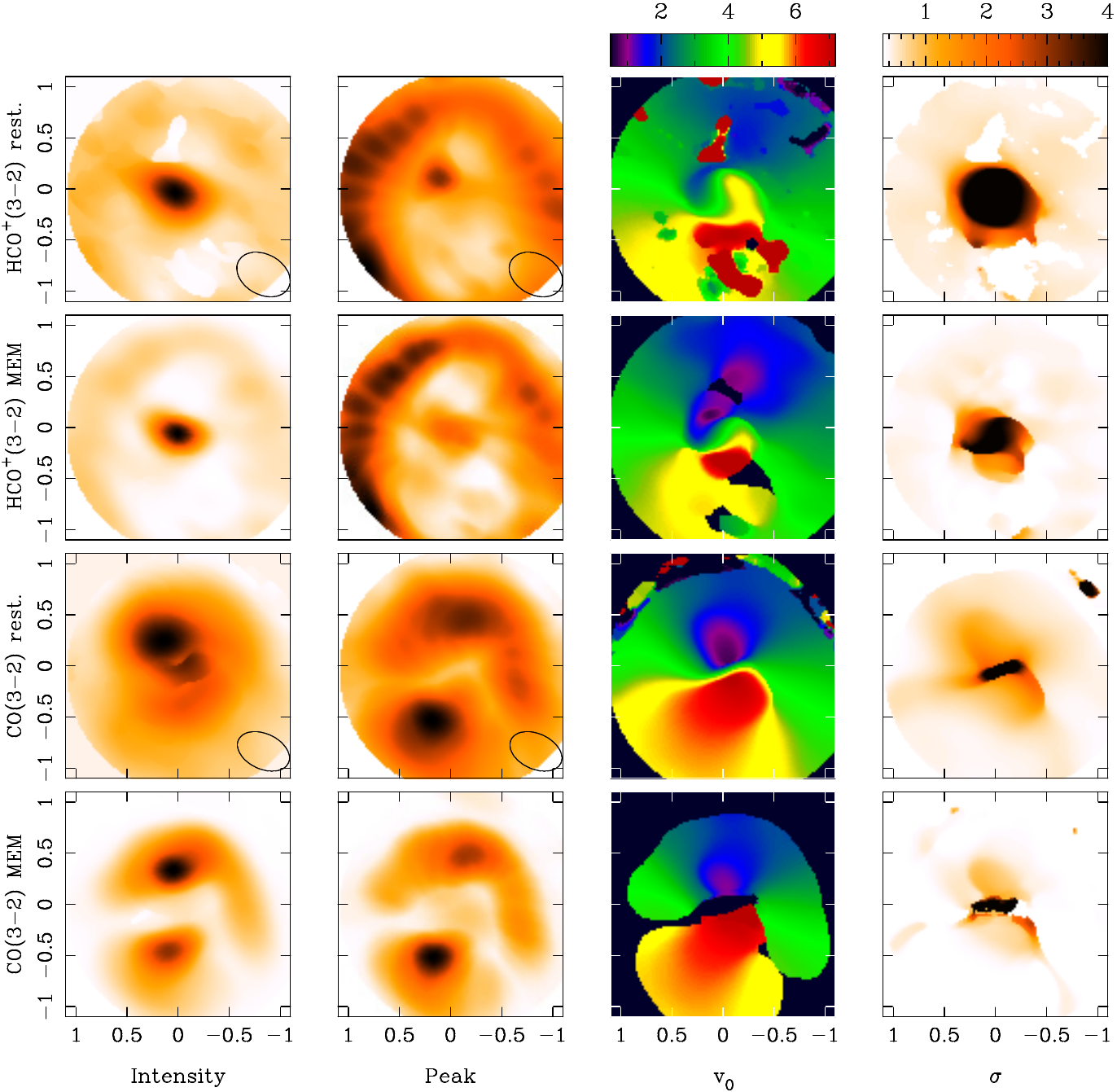}
\end{center}
\caption{Moment maps in HCO$^+$(4-3) and CO(3-2). For each specie, we
  plot both restored images (labeled {\tt rest.}) and the
  corresponding deconvolved models (labeled {\tt MEM}). Columns
  correspond to spectral moments, from left to right we show:
  intensity (see Appendix~\ref{sec:momentmaps}), specific intensity
  peak, Gaussian velocity centroid, and Gaussian 1~$\sigma$ velocity
  width.
 \label{fig:comple_moments}}
\end{figure*}

Fig.~\ref{fig:comple_moments} provides a summary of the data in each
transition. It is apparent that the coarser beam in band~7, compared
to band~9, prevents a detailed study of the kinematics in the central
0.3\arcsec. We also see in the deconvolved maps that attempts to trace
the CO(3-2) kinematics inside the cavity are thwarted by the
absorption from the intervening diffuse cloud in the line of sight to
HD~142527 \citep[e.g.][]{Casassus2013A&A...553A..64C}.  Once convolved
with the clean beam the regions affected by the intervening CO cloud
are less conspicuous, and may lead to an accidental interpolation of
the velocity field \citep[as in][]{Rosenfeld2014ApJ...782...62R}. Due
to this intervening absorption, the PA in CO(3-2) as a function of
radius does not appear to rotate beyond 30--40\degr\, East of North, or
$\lesssim$60\degr\, relative to the outer disk. This shift is close to the
value of 54\degr\, obtained when the radial velocity component scales as
$\sqrt{2}$ times the azimuthal velocity component (i.e. for a radial
component close to free-fall).

\section{Modeling} \label{sec:model}

\subsection{Parametric modeling} \label{subsec:radmc3d}

In order to reproduce the intra-cavity kinematics, we developed on the
accretion flows proposed to account for the HCO$^+$ observations
\citep{Casassus2013Natur}, while also taking into account the warp
implied by the scattered-light shadows
\citep{Marino2015ApJ...798L..44M}. We used a parametric model for
HD~142527 that fits the spectral energy distribution \citep[see][for a
  detailed description]{Marino2015ApJ...798L..44M}, with the addition
of a radial velocity component corresponding to stellocentric
accretion.  Emergent specific intensity fields were predicted by
transferring the stellar radiation through the model circumstellar
material with the {\sc
  radmc3d}\footnote{http://www.ita.uni-heidelberg.de/
  dullemond/software/radmc-3d/} package \citep[version
  0.39,][]{RADMC3D0.39}.

The warp is implemented by rotating the disk as a function of
stellocentric radius.  The orientation of the outer disk, with a fixed
inclination $i=20$\degr\, and PA=$-20$\degr\, beyond $r_{\rm{orient\_out}} = 23$~AU, is linked
linearly at $r_{\rm{orient\_in}} = 20$~AU with a non-coplanar inner disk, at a relative
inclination of 70\degr\, and PA$=-8$\degr. The orientation of the inner
disk is such that its inclination relative to the plane of the sky is
$i\sim 46.7$\degr, so that the Eastern side is near, while East is far
for the outer disk\footnote{in other words the parallactic angle of
  the normal to the disks is rotated by 168\degr}. The inner disk
probably contains the orbit of the low mass companion reported by
\citet{Biller2012} and \citet{Close2014ApJ...781L..30C}.



\begin{figure}
\begin{center}
\includegraphics[width=0.9\columnwidth,height=!]{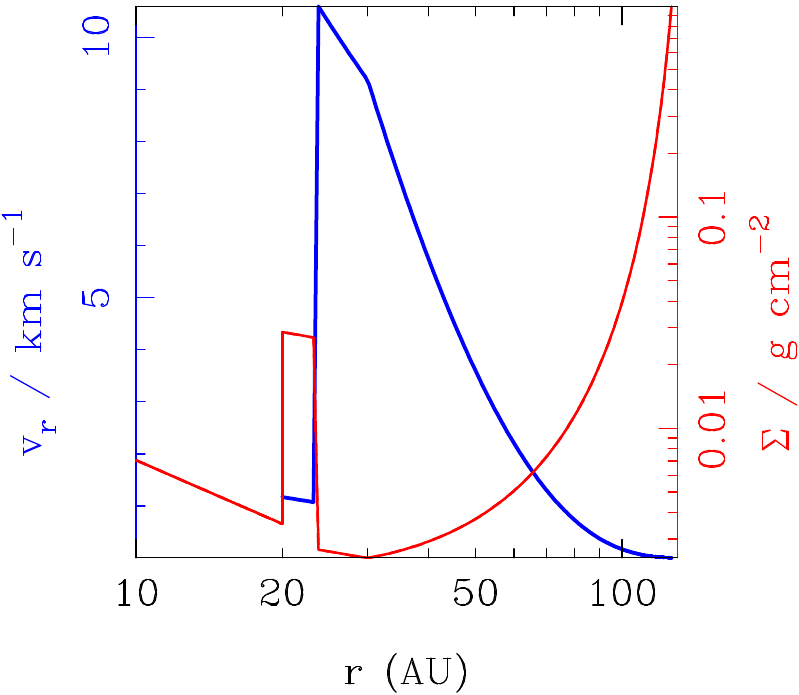}
\end{center}
\caption{Radial cuts for the surface density profiles (red) and
  magnitude of the radial velocity component (blue).  The CO abundance
  is modulated with a Gaussian taper inside 30~AU, to account for
  photo-dissociation.  \label{fig:cavity_profiles}}
\end{figure}

Radial cuts illustrating our choice of velocity and surface density
profiles are shown in Fig.~\ref{fig:cavity_profiles} inside the
cavity.  We caution that this parametric model is meant to help the
interpretation of the data, and is not derived from a systematic
search in parameter space. It is not a dynamically consistent model.
Along with the warp, we also add a radial velocity component that
accounts for accretion. We connect zero accretion at $\sim$140~AU to
free-fall at 30~AU, with a cubic profile.  The free fall halts at
23.3~AU. The tilted inner disk is assumed to be in pure Keplerian
rotation, so that inside 20~AU we set zero accretion (a build-up of
accreted mass onto the inner disk could then trickle to feed pole-on
stellar accretion). The total surface density profile in the gap follows from
the assumption of steady state accretion:
\[
\Sigma(r) = \frac{dM_\star}{dt} / (2 \pi r v_r), 
\]
as a function of spherical polar radius $r$, where $v_r$ is the
radial velocity component,
\begin{eqnarray}
&v_r = 0   ~\rm{if} ~   r < r_{g1},   ~ \rm{with} ~r_{g1} = 20~\rm{AU}, ~\rm{or}~  & \nonumber \\ 
&v_r = -0.1v_{\rm{ff}}   ~\rm{if}~   r < r_{\rm{stop}},  ~ \rm{with}~ r_{\rm{stop}} = 23.3~\rm{AU}, ~\rm{or}~ &  \nonumber \\ 
&v_r = -0.9 v_{\rm{ff}}   ~\rm{if}~  r < r_{\rm{ff}},  ~ \rm{with}~ r_{\rm{ff}}= 30~\rm{AU}, ~\rm{or}~ &   \nonumber \\
&v_r = -0.9 v_{\rm{ff}}  \times  \left[ \frac{ (f\,r_{g2} - r ) }{ (f\,r_{g2} - r_{g1})} \right]^3   ~\rm{if}~  r < r_{g2},   & 
\end{eqnarray}
with $ r_{g2} = 130~\rm{AU}$, and where $f=1.1$ is a factor to avoid divergence at $r_{g2}$, and
\[
v_{\rm{ff}} = \sqrt{  2 G M_\star \times \left( \frac{1}{r} - \frac{1}{f\,r_{g2}} \right) }.  
\]
We use a stellar accretion rate $dM_\star/dt = 10^{-7}~M_\odot$
\citep{2006A&A...459..837G}. The stellar mass is set at $1.8~M_\odot$;
higher masses produce higher velocities inside the cavity than
observed.

The orientation of the disk is parametrised in spherical coordinates,
where the vertical axis $\hat{z}^\prime$ is the normal to the local
disk plane, and where the origin of azimuth $\phi^\prime$ is the local
line of nodes (i.e. the disk PA as a function $\vec{r}$).  The
azimuthal velocity $v_{\phi^\prime}$ is set to Keplerian rotation $v_{\phi^\prime} =
v_K$, except in regions where $|v_r| > 0$, where we assumed that the
specific energy density was close to 0, so that $v_\phi =
\sqrt{v_{\rm{ff}}^2 - v_r^2}$ (neglecting the internal energy).

In the warp itself, so between $r_{\rm{orient\_in}} = 20$~AU and
$r_{\rm{orient\_out}}=23~$AU, material continuously connects both
orientations. We investigated a range of values for the vertical
velocity component in the warp, and obtained a closer match to the
observations with relatively fast velocities, $v_{\theta^\prime} =
v_{\rm warp} = v_K \sin(\phi^\prime)$ (see
Sec.~\ref{subsec:centralbeam}).

The temperature profile we adopted is $T=70~$K if $r < 30$~AU, or 
\begin{equation}
  T(r) = 70 \left(\frac{r}{30 {\rm AU}} \right)^{-0.5} ~{\rm K}, ~\rm{if}~ r>30~\rm{AU}.
\label{eq:temp}
\end{equation}
These temperatures are well below the dust temperatures in the
closer-in regions. If we impose thermal equilibrium between both solid
and gas phases, the predicted flux densities soar an order of
magnitude above the observations.


Close to the star CO is probably photo-dissociated. The line intensity
of CO(6-5) appears to display a decrement inside $0.2$\arcsec, which
we accounted for with a Gaussian taper of the gas abundance inside
30~AU, with a $1~\sigma$ width of $5$~AU.

\subsection{Radiative transfer predictions} \label{subsec:modelresults}

\subsubsection{Adopted abundance fields and depth of the cavity}

We adopted a nominal H$_2$/$^{12}$CO ratio $X_{\mathrm{CO}}=10^{-4}$,
and the following molecular abundances relative to H$_2$ (no selective
photodissociation was required). For $^{13}$CO we used the ISM
isotopic abundance for $^{12}$C/$^{13}$C of 76
\citep{2008A&A...477..865S}, multiplied by $X_{\mathrm{CO}}$. For
C$^{18}$O, we proceeded as for $^{13}$CO, but with an ISM isotopologue
abundance for C$^{18}$O/$^{12}$CO of 500
\citep{1994ARA&A..32..191W}. For HCO$^+$, we used a default value for
the HCO$^+$ abundance relative to $H_2$ of $10^{-9}$ \citep[somewhat
  lower than observed in molecular cloud cores, and comparable to
  values inferred in other disks,][]{Wootten1979ApJ...234..876W,
  Teague2015A&A...574A.137T}.

%

%

\begin{figure}
\begin{center}
\includegraphics[width=\columnwidth,height=!]{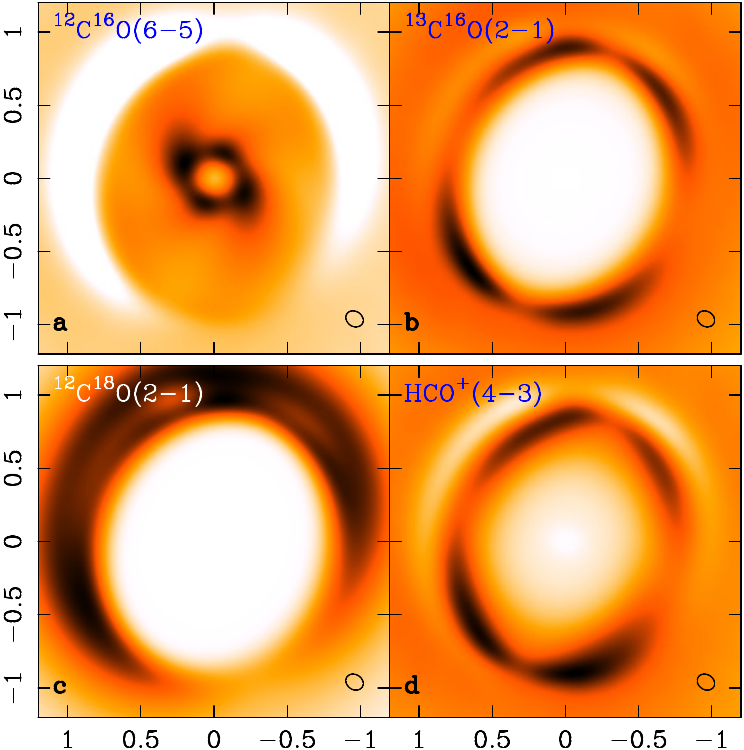}
\end{center}
\caption{Intensity maps (moments 0)  extracted from the RADMC3D predictions, after
  smoothing to the resolution of the deconvolved CO(6-5)
  datacubes. \label{fig:co65moments_radmc_smooth}}
\end{figure}

After smoothing to the resolution of the deconvolved models,
Fig.~\ref{fig:co65moments_radmc_smooth} shows that the accreting
warped disk model qualitatively reproduces the basic properties of the
cavity seen in the available molecular tracers.  The model is in
agreement with the non-detection of $^{13}$CO and C$^{18}$O inside the
gap. Both in band~6 \citep{Perez2015ApJ...798...85P} and band~7
\citep{2013PASJ...65L..14F}, the $^{13}$CO and C$^{18}$O datacubes do
not show any sign of either twisted kinematics, a central peak, or of
the gap-crossing filaments. The CO isotopologue data display plain
Keplerian rotation, without signal inside $\sim$90~AU.

The decrements under the crescent-shaped continuum, previously
reported by \citet{Casassus2013Natur}, are a natural feature of the
predictions. In continuum-subtracted images, and with our gas
temperature prescription, the gas effectively acts as a cool
foreground \citep{Casassus2015hd142trap}.


%

%
%

\subsubsection{Central beam}  \label{subsec:centralbeam}

\begin{figure*}
\begin{center}
  \includegraphics[width=\textwidth,height=!]{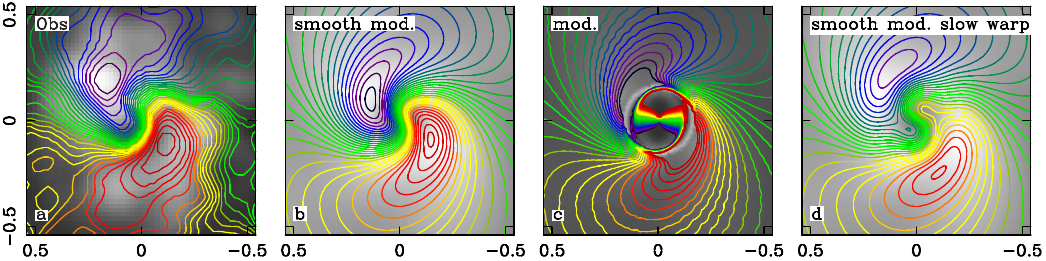}
\end{center}
\caption{Comparison of observed and model CO(6-5) kinematics in the
  central regions of HD~142527. The origin of coordinates is set to
  the stellar position.  Velocity-integrated intensity in CO(6-5) is
  shown in grey scale. The colored contours for $v_\mathrm{lsr}$ have
  constant interval and are spread over $[0.21,7.87]$~km~s$^{-1}$ (as
  in Fig.~\ref{fig:co65moments_memonly}).  {\bf a)}: Observed moment
  maps extracted on the MEM datacubes. {\bf b)}: moments extracted on
  the radiative transfer prediction, after smoothing to the resolution
  of the MEM datacubes. {\bf c)}: same as b) but in native model
  resolutions, without smoothing. Regions without contours near the
  origin correspond to higher velocities.  {\bf d)}: same as b) but
  with a slow velocity component perpendicular to the disk plane
  ($v_{\rm{warp}}$ in the text).
  \label{fig:zoom_obsmod} }
\end{figure*}

The warp inside the cavity of HD~142527, i.e. the continuous change in
inclination at some specific radius, should significantly impact the
observed kinematics. For instance, given the orientation implied by
the scattered light shadows \citep{Marino2015ApJ...798L..44M}, the
blue-shifted flows to the North should suddenly shift to red as the
inclination changes sign, at the radius where the far side becomes the
near side. Interestingly, as illustrated in finer detail in
Fig.~\ref{fig:zoom_obsmod}a, the observed centroid map indeed has
structure very close to the central star, well inside the central
beam. Within $\sim$0.25\arcsec, the `twist' in PA, i.e. the rotation
of the axis joining blue and red-shifted velocities, appears to dwell
at $\sim$40\degr, and so offset by $\sim$60\degr\, from the outer disk
PA, as expected for a radial velocity component that has reached close
to free-fall values while also keeping in near-Keplerian
rotation. Indeed, at a given stellocentric radius, in the presence of
a radial velocity component that is $\sqrt{2}$ times the azimuthal
component, the peak line-of-sight velocity is offset from the disk PA
by $\sim$54.6\degr. However, inside 0.1\arcsec, the PA joining blue
and red shifted velocities appears to reach $110\pm10$\degr.

The extreme in PA rotation in the inner 0.1\arcsec~ cannot be
accounted for by free-falling material that is also in close to Keplerian
rotation. This can instead be understood as the halt of free-fall
inside a warped structure, as the blue-shifted flows to the North-West
(11h) suddenly doppler-flip into red closer to the star, where the
inclination changes sign (far side turns near).  After smoothing to
the resolution of the deconvolved MEM datacubes, we obtain an
approximate match to the velocity field for an abrupt warp connecting
the inner disk at 20~AU with the orientation of the outer disk at
23~AU, as shown in Fig.~\ref{fig:zoom_obsmod}b. The warp orientation
must align to that of the inner disk as far out as 20~AU, to account
for the shift in declination at systemic velocities, which covers
almost 0.2\arcsec~ when crossing the stellar position.

In a continuous warp material must flow from one plane to the other
and be accelerated in the transition. So inside a warp there should be
a velocity component orthogonal to the plane of the disk $v_{\rm
  warp}$.  This component is probably limited by the Keplerian
velocity $v_K$ around the central star. For instance with $v_r =
0.1~v_{\rm ff}$ and $v_\theta = v_{\rm warp} = v_K$ in spherical
coordinates, the warp's radial width must be at least $\sim$3AU to
connect one orientation to the other. Interestingly, the inclusion of
$v_{\rm warp}$ approximates better the observed velocity centroid,
with fairly point-symmetric high velocity ridges connecting the cavity
`twist' with the inner disk, corresponding to the loci of bluest and
redest velocities in Fig.~\ref{fig:zoom_obsmod}~b.  These ridges are
missing in the model without $v_{\rm warp}$
(Fig.~\ref{fig:zoom_obsmod}d).



\section{Discussion} \label{sec:discussion}

\subsection{Ro-vibrational CO} 

The CRIRES observations by \citet{Pontoppidan2011ApJ...733...84P}
assign bright rovib CO to an inner disk, with similar inclination as
the outer disk but offset in PA, at $61 \pm 3$\degr\, - so within
uncertainties coincident with the direction of the high-velocity
HCO$^+$. The total line-widths in CO($v$=1-0) at 4.67$\mu$m, at
zero-intensity, are close to $\sim$40~km~s$^{-1}$
\citep{Pontoppidan2011ApJ...733...84P}. With a spectroastrometric
offset of only 0.2~AU, CO($v$=1-0) is either fairly uniform or else
confined to the central few milli-arcsec.

We check the astrometric signal of a modified version of our disk
model and find it to be in agreement with the spectro-astrometric
CRIRES observations. We take the predicted CO(6-5) emission and
transform the intensity with an $1/r^\alpha$ mask to mimic the CO
ro-vibrational emission, which traces higher temperatures and emits
closer to the star compared to the CO rotational lines. We manually
add continuum emission scaled to reproduce a line to continuum ratio
of 1.1, as in the observed rovibrational CO, and convolve with a
0.192" FWHM Gaussian to mimic the PSF of the AO-assisted CRIRES
observations. After superimposing a 0.2" arcsecond wide slit
positioned at 150 degrees, we recover a line profile and
spectro-astrometric signal similar in shape and magnitude to that
reported by \citet{Pontoppidan2011ApJ...733...84P} when using $\alpha
\sim 2.5$.

%

%
%

\subsection{What about an outflow?}

The orientation of the disk is well known.  As explained in
Secs.~\ref{sec:intro} and Sec.~\ref{subsec:radmc3d}, the shadows seen
in scattered light imply that the inclination of the outer disk is
such that the far side lies to the East. The Eastern side being
brighter in the mid-IR is also in support of this orientation
\citep{2006ApJ...644L.133F}.  In addition several near-IR spirals have
been found that are all consistently trailing \citep{Fukagawa2006,
  Casassus2012ApJ...754L..31C, Canovas2013A&A...556A.123C,
  Avenhaus2014ApJ...781...87A}, even in their radio extensions into
the outer disk \citep{Christiaens2014ApJ...785L..12C}. Given this orientation, we can
rule out an outflowing disk wind: the velocity component orthogonal to
the disk would broaden the lines and preserve reflection symmetry
about the outer disk PA, while the radial component in the plane of
the disk would twist the kinematics in the opposite direction than
observed.

\subsection{Origin of the free-falling velocities}

The radial velocity in a warped viscous disk is predicted to be
greater than in a flat disk \citep{Pringle1992MNRAS.258..811P}, but is
not expected to exceed the sound speed $c_s = (h/r) v_K$. Supersonic
disk inflow is in principle possible in regions of the disk that lose
angular momentum rapidly through a magnetized wind, rather than via
internal viscous transport, though whether this can occur without
equally rapid mass loss (for which there is no evidence in HD~142527)
is not clear. Alternatively, the free-fall velocity signal could be a
consequence of strong dynamical perturbations. If the low-mass
companion detected in the system \citep{Biller2012,
  Close2014ApJ...781L..30C} is misaligned to the outer disk in the
same way as the inner disk, which is known to be inclined by 70\degr\,
\citep{Marino2015ApJ...798L..44M}, then an observable dynamical impact
is highly likely. The alignment of the companion in the plane of the
inner disk seems a natural way to explain the inclination of the inner
disk itself. \citet{Miranda_Lai_2015arXiv150402917M} have recently
shown that large offsets between the orbital planes of circumprimary
disks and the plane of the binary can be stable for several Myrs,
longer than the expected lifetime of the disk.

%

It has been shown recently that such strongly misaligned binary-disk
systems can undergo a process dubbed disk tearing
\citep[][]{Nixon_2013MNRAS.434.1946N,
  2015MNRAS.448.1526N,2015MNRAS.449.1251D}, where nodal precession
torques induced by the binary produce a warp at the inner edge of the
disk. If the disk is unable to communicate the warp efficiently, then
the disk will effectively break at some radius $R_{\rm break}$. Annuli of
gas then begin to tear off of the inner edge and can precess freely,
undergoing self-interaction and angular momentum cancellation causing
a subsequent infall of gas driving very high accretion rates
\citep{2012ApJ...757L..24N}.

To date this process has primarily been studied in the thin-disk
regime appropriate for black hole accretion, where the dimensionless
viscosity parameter is larger than the disk aspect ratio, i.e. $\alpha
> H/R$ \citep[][]{2012ApJ...757L..24N,
  Nixon_2013MNRAS.434.1946N}. Nevertheless it has also been shown to
work in the thick-disk regime where $\alpha < H/R$
\citep{2015MNRAS.448.1526N}, which is more likely to hold true for
circumstellar disks like HD~142527.

It is possible to estimate a value for the radius $R_{\rm break}$ at
which disk tearing will occur \citep[see Eq. A3
  of][]{Nixon_2013MNRAS.434.1946N} using the parameters of the disk
and binary. With appropriate values for HD~142527, we expect that the
disk could break at $R_{\rm break} \sim a$, where $a$ is the binary
semimajor axis. However, \citet{2015MNRAS.448.1526N} found a
consistent offset between the estimated minimum value and the true
breaking radius in a series of SPH simulations, by a factor of
2--3. Estimates of $a \sim15$~AU are therefore entirely consistent
with a breaking radius of 30~AU, as demanded by our model. However, as
this is likely very sensitive to the parameters of the system a more
quantitative comparison of this scenario with the observed data is out
of the scope of this paper, requiring the development of targeted
numerical simulations.

We also note that while this scenario seems to adequately explain both
the inner warp and the free-falling velocities, it does not explain
the presence of the $\sim130$~AU cavity in the system. This should not
be surprising, given the large difference in scale between the warp
and the cavity (and therefore different timescales associated with
physical processes at these radii), but means that the model is not
yet a complete description of the system.

\subsection{Long term Kozai oscilations and the large cavity} \label{sec:kozai}

The dynamical influence of the low-mass companion on its current orbit
will not extend out to the 100 AU scale needed to explain the large
extent of the cavity. It is possible, however, that its orbit may have
changed over time. The highly inclined orbit of the companion relative
to that of the massive outer disk could trigger Kozai oscillations
\citep{Teyssandier2013MNRAS.428..658T}, with a period of the order of
$10^5$~yr if the outer disk has a mass of 0.1~M$_\odot$. A Kozai cycle
would result in (damped) oscillations between the current inclined
circular orbit, and a highly eccentric coplanar orbit which would
perturb the disk out to greater radii. Numerical simulations are
needed to determine if such a cycle is possible in HD
142527.

\subsection{Non-axial symmetry in the cavity. Signature of obscured planets?} \label{sec:filaments}

Observationally we cannot test directly for past coplanar eccentric
orbits, as in the Kozai oscilations, but we can instead look for other
mechanisms that could account for the clearing, such as additional
bodies at $\sim$100~AU. Stringent limits are available
\citep{Casassus2013Natur}, $\sim$4~$M_\mathrm{jup}$ bodies should have
been detected unless they are obscured.

%

\begin{figure}
\centering
\includegraphics[width=0.9\columnwidth]{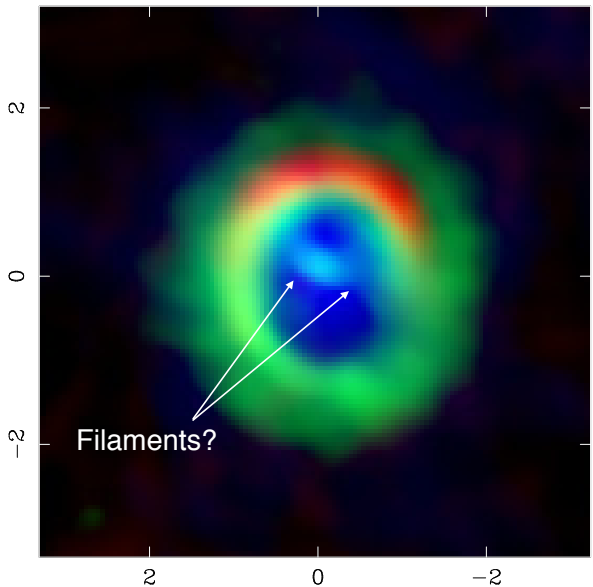}
  \caption{Summary of Cycle~0 band~7 observations, from MEM maps, with
    continuum in red, HCO$^+$(4-3) in green, and CO(3-2) in
    blue. Velocities have been restricted to highlight the fainter
    structures seen in HCO$^+$, which are otherwise dwarfed by the
    fast HCO$^+$ central emission. 
  \label{fig:ALMARGB}}
\end{figure}

An interesting feature of the Cycle~0 HCO$^+$(4-3) data is the faint
filament crossing the cavity to the East, at low velocites, with
perhaps another filament to the West \citep[see Fig.~\ref{fig:ALMARGB}
  for a summary,][]{Casassus2013Natur}. It is tempting to associate
these HCO$^+$ features with planet-induced accretion flows,
i.e. gaseous streamers due to obscured protoplanets inside the cavity,
that feed the faster flows at the disk breaking radius. As the
critical density for the excitation of HCO$^+$(4-3) is $n_\mathrm{H2}
\sim 10^6$~cm$^{-3}$, and higher than the CO(3-2) critical density of
$\sim 10^{3}$~cm$^{-3}$, in the streamer interpretation the HCO$^+$
filaments and high-velocity flows would trace the densest regions of
the general accretion flow across the gap, which is also reflected in
the `twisted' CO kinematics. If the free-falling accretion is
initially fed by planet accretion at $\sim~100$AU, there should be
corresponding azimuthal modulations of the H$_2$ density
field. Perhaps this is reflected in the structure of the faint,
low-velocity HCO$^+$(4-3) inside the cavity.

An alternative to azimuthal modulations of the H$_2$ density field in
the cavity, is that only the abundance of HCO$^+$ is modulated. The
tilted inner disk is shadowing the UV radiation required to produce
HCO$^+$ in the cavity, resulting in HCO$^+$ filaments with wide
opening angles, as opposed to the thin protoplanetary streamers
predicted in hydrodynamic simulations. A problem with this
interpretation is that the outer ring is fairly round in HCO$^+$,
there are no obvious counterparts in HCO$^+$ of the scattered-light
shadows. Perhaps this reflects a different production chain for
HCO$^+$ in the outer disk, where it could be driven by charge-exchange
with cosmic-ray induced H$_3^+$.

\section{Conclusion} \label{sec:conclusion}

The new CO(6-5) data, along with the orientation of the disk inferred
from the scattered light shadows, have allowed us to understand the
intra-cavity kinematics in HD~142527.  Stellocentric accretion,
starting from the outer disk and reaching close to free-fall
velocities, with a steady-state mass flow fixed at the observed
stellar accretion, is consistent with the bulk properties of the
available CO isotopologue data. The observed HCO$^+$ flows are also
consistent with this stellocentric accretion, but with an emissivity
that is somehow modulated in azimuth. Fine structure in CO(6-5) also
suggests non-axial symmetry inside the cavity.

While the data are consistent with a continous, yet abrupt and fast
warp linking the two non-coplanar disks, further observations are
required to understand the detailed structure of the intra-cavity
kinematics and the inner warp. For instance, the large relative
inclination observed between the inner and outer disks suggests that
the fast accretion could be due to disk tearing. If the low-mass
companion is contained in the inner disk and is breaking the disks,
then its orbit would also be highly inclined with respect to the outer
massive disk. The companion may then undergo Kozai oscillations, with
high-eccentricity periods that may perhaps explain the large cavity.

%

%

%

\acknowledgments

Based on observations acquired at the ALMA Observatory, through
program {\tt \small \sc ALMA\#2011.0.00465.S}. ALMA is a partnership
of ESO, NSF, NINS, NRC, NSC, ASIAA. The Joint ALMA Observatory is
operated by ESO, AUI/NRAO and NAOJ. Financial support was provided by
Millennium Nucleus RC130007 (Chilean Ministry of Economy), and
additionally by FONDECYT grants 1130949, 1141175, 3140601, 3140634,
3140393. S.M. acknowledges CONICYT-PCHA / Magister Nacional /
2014-22140628.  PR and VM acknowledge support from CONICYT-ALMA grant
ALMA-CONICYT 31120006. AD acknowledges CONICYT-ALMA grant 31120007.
PJA acknowledges support from NSF award AST 1313021. MM acknowledges
CONICYT-Gemini grant 32130007. This work was partially supported by
the Chilean supercomputing infrastructure of the NLHPC (ECM-02).

\appendix

\section{Observations}


\subsection{ALMA CO(6-5) data} \label{sec:data}

Details on the instrumental setup are described in a companion article
on the continuum emission \citep{Casassus2015hd142trap}. We used
self-calibration to improve the dynamic range of the continuum images.
Applying the same gain corrections to the line data also resulted in
improved dynamic range. Before self calibration, the peak signal in
the systemic velocity channel reached 0.42~Jy~beam$^{-1}$, in natural
weights (beam of $0.30\arcsec\times0.25\arcsec$), with a noise level
of 0.08~Jy~beam$^{-1}$ clearly dominated by systematics rather than
thermal noise. After self calibration, the peak signal increased to
0.67~Jy~beam$^{-1}$, while the rms noise level decreased to
0.04~Jy~beam$^{-1}$.

Continuum subtraction under CO(6-5) was performed with a first-order
fit to the continuum in the visibility domain. The subset of channels
neighboring the line was then split off into another datafile, and
subsequently re-sampled in frequency into the LSRK frame.

\begin{figure}
\begin{center}
\includegraphics[width=!,height=0.85\textheight, angle=-180]{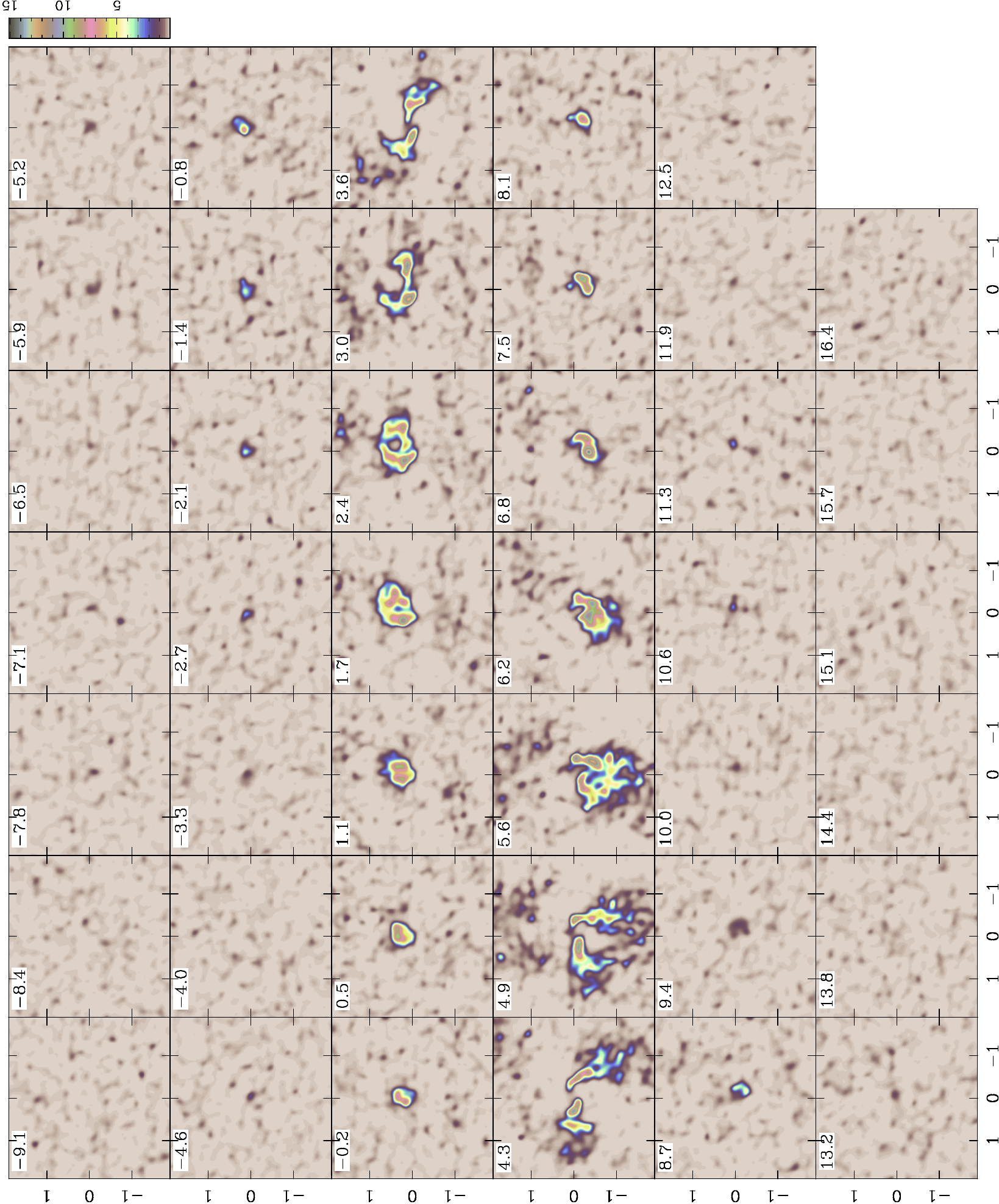}
\end{center}
\caption{Channel maps in the {\tt uvmem} model datacube for
  CO(6-5). We binned the datacube into 3-channel averages.}\label{fig:channels_model}
\end{figure}

\begin{figure}
\begin{center}
\includegraphics[width=!,height=0.85\textheight, angle=-180]{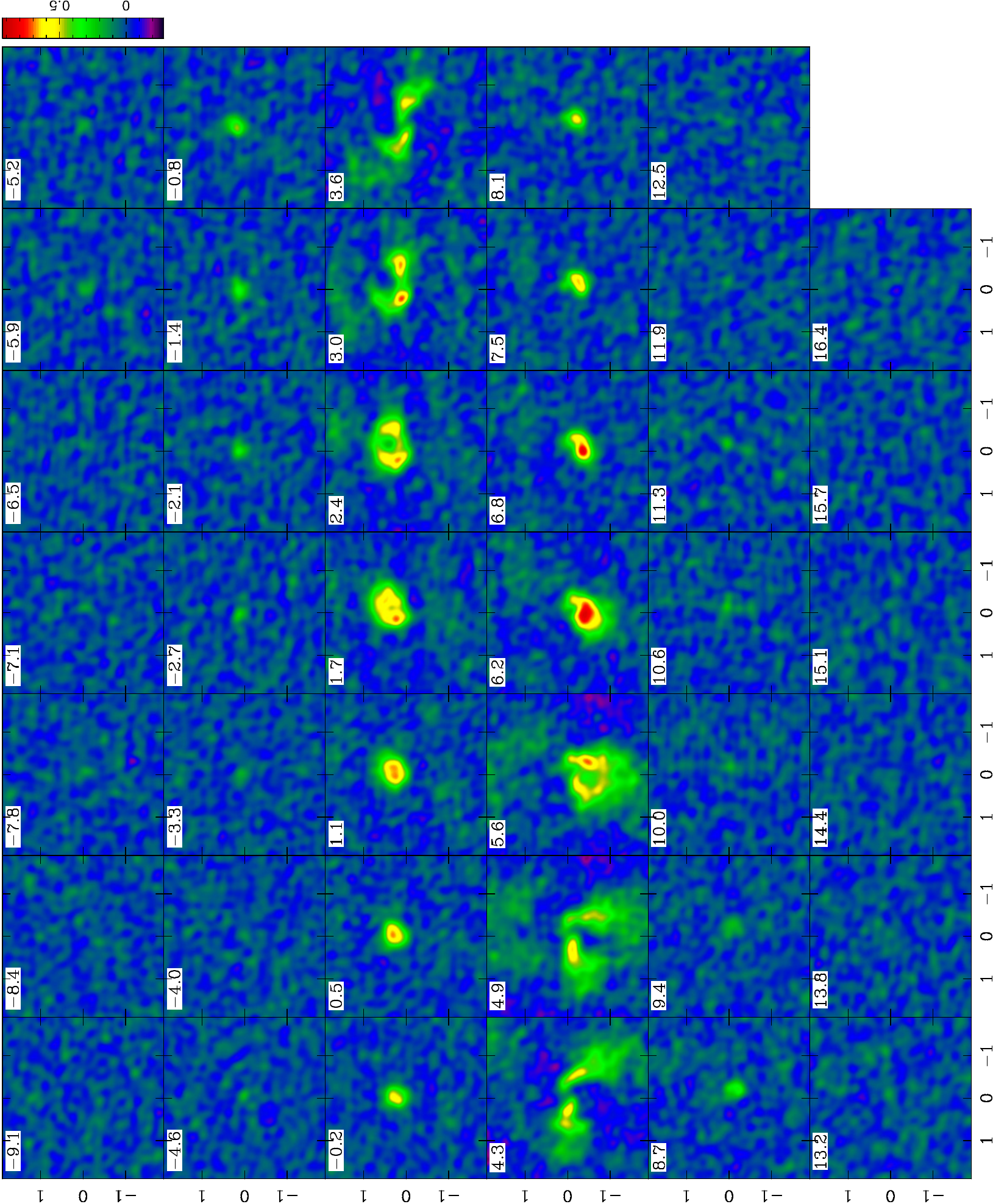}
\end{center}
\caption{Channel maps in the restored CO(6-5) datacube. We binned the
  datacube into 3-channel averages.
}\label{fig:channels_restored}
\end{figure}

The CO(6-5) dataset is presented in channel maps in
Figs.~\ref{fig:channels_model} and ~\ref{fig:channels_restored}, with
three-channel bins. We chose to present both the restored image
and the underlying MEM model. 

\subsection{Alignment}

%
%

The comparison of the band~9 and band~7 continua reveals a positional
offset of $\sim$0.07~arcsec essentially in the East-West direction
(precisely $[-0.063, -0.016]$ in arcsec, along RA and DEC). As
detailed in a companion article devoted to a multi-frequency analysis
of the continuum, including a detailed presentation of the band~9
continuum, this positional offset is measured by cross-correlating the
visibilities, but it is also noticeable by visual inspection of the
continuum maps (for instance by blinking the images).  With the
0.35~arcsec resolution provided by these Cycle~0 observations in
band~7, the center of symmetry of the CO(3-2) velocity centroids
coincide fairly well with the position of the central star at the
epoch of observations. Hence for the multi-frequency continuum studies
we chose to take the band~7 astrometry as reference.

However, at 0.02~arcsec resolutions the center of symmetry of the
velocity centroid appears to be offset from the nominal stellar
position as given by the default band~7 world-coordinate-system
(WCS). This represents an opportunity to improve the astrometry by
tying the stellar position with the center of symmetry. We aligned the
velocity centroids by applying an offset of $ [ -0.109, -0.031
]$~arcsec to the default band~9 WCS. As illustrated in
Figs.~\ref{fig:co65moments_memonly}, \ref{fig:zoom_obsmod} and
\ref{fig:co65moments_compar}, the center of symmetry of the velocity
centroid coincides with the origin of coordinates, located at the
stellar position. The peak velocity dispersion is also close to the
origin, but in this case the exact location of the peak depends on the
moment extraction algorithm, as the signal is spread over a large
range in velocity and includes low signal-to-noise channels.

%
%

\section{Image synthesis} \label{sec:imagesynthesis}

\subsection{Angular resolution and statistics of the deconvolved MEM models}
  
Details on our image synthesis algorithm {\tt uvmem} are given in a
companion paper on the continuum emission
\citep{Casassus2015hd142trap}. Here we exploit the finer angular
resolutions provided by the MEM models, which requires an assessment
of their statistical properties.

We start by scaling the observed visibility weights so that they
correspond to the observed scatter. In order to preserve the relative
values of the calibrated weights, a constant scale factor is applied
to all visibilities. This factor is the median of the ratio between
the original weights and the dispersion values.

Estimates of the angular resolution of the deconvolved images were
obtained as follows.  We selected a reference channel, at the systemic
velocity, and split off the corresponding visibilities into a
reference visibility dataset. We then estimated the PSF by simulating
interferometer observations on a point source, with an identical $u,v$
coverage as the reference visibilities, and with the addition of
Gaussian noise, with a dispersion given by the visibility weights.  In
practice the point source is a spike, defined on the same canvas as
used for the deconvolved images, and whose flux density corresponds to
that of the whole observed image. A deconvolved model of these
simulated point-source visibilities is then obtained using the same
algorithm as for the observations (i.e. {\tt uvmem}). This deconvolved
model represents the PSF of the deconvolution algorithm, with a
central source and secondary maxima fainter by a factor of at least
20. An elliptical Gaussian fit to the central peak defines the finest
resolution achievable in the deconvolved images, of 0.062\arcsec~
$\times$ 0.051\arcsec.

Since the maximum-entropy regularization is non-linear, the {\tt
  uvmem} PSF depends on the model. We approximate the case of extended
emission by spreading the total flux in a collection of spikes. The
`clean' beam in this case varies by up to 50\%. So, for the purpose of
comparing synthetic radiative transfer predictions with the {\tt
  uvmem} model images, we chose a representative resolution of 30\%
times coarser than the finest PSF.

Monte-Carlo simulations provide estimates of the uncertainties
associated with these MEM model datacubes.  We added Gaussian noise to
the observed visibilities, as given by the visibility weights, and
obtained a deconvolved model image with {\tt uvmem}, with the same
settings as for the observations. We then took statistics on 120
realizations, checking that the average datacube satisfactorily
converges on the original model datacube without the addition of
noise.

\subsection{Moment maps} \label{sec:momentmaps}

In the case of noisy data, the standard approach to extract moment
maps by collapsing the velocity axis injects unnecessary noise when
the line-of-sight line profiles are much narrower than the spatially
integrated line.  Rather than collapse the datacubes with standard
moments, we extracted moments by restricting the velocity range above
a given threshold (which is set to 0 for the restored maps, with
thermal noise).  For each line of sight, we identify the position of
the peak specific intensity. We then search for the first nulls, both
towards the red and blue, and extract a spectrum leaving room for a
few more channels (two in this case). Last, we take standard moments
over this reduced velocity range, and also fit a Gaussian. The
resulting moment maps are embellished by median-smoothing with a
square $5\times5$ kernel.

In the case of the MEM deconvolved models, the strict positivity
requirement forces an intensity floor $I_{\rm min}$, which we take at
1/100 the estimate noise level (in natural weights). We subtracted
{\em a-posteriori} $I_{\rm min}$ from the deconvolved maps. However,
the positivity requirement biases the noise, so that velocity
dispersions are artificially broad in region of low signal (such as
the outer ring in HD~142527 in CO(6-5)). To account for this we applied the
velocity restrictions described above, with a threshold defined by the
rms dispersion of the noise in velocity channels devoid of signal.

\subsection{Comparison of different image synthesis schemes} \label{sec:synthesis_compar}

\begin{figure*}
\begin{center}
\includegraphics[width=0.6\textwidth,height=!]{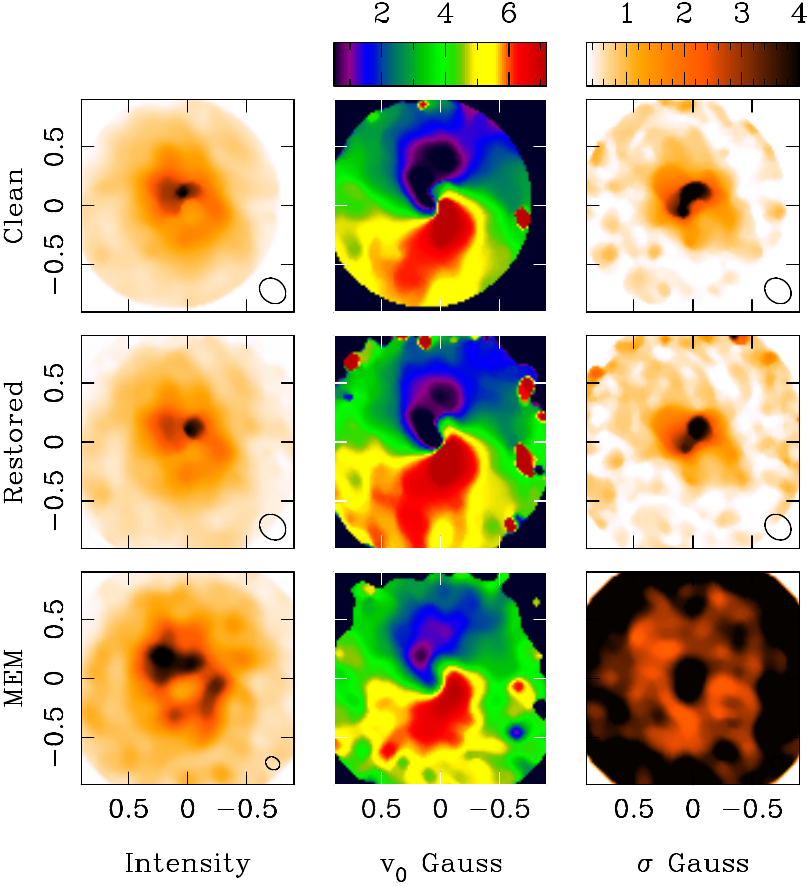}
\end{center}
\caption{Moment maps in CO(6-5) emission. $x-$ and $y-$axis show
  angular offset along RA and DEC from the stellar position. The upper
  row shows maps extracted on the {\tt clean} datacubes, produced in
  Briggs weights. The middle and bottom row respectively show the maps
  extracted on the restored {\tt uvmem+CASA} datacubes, in Briggs
  weights, and the deconvolved MEM models. Columns correspond to
  intensity (left), velocity centroid (middle), and 1~$\sigma$
  velocity dispersion (right). The moments from the deconvolved MEM
  maps are extracted without the thresholds
  (Sec.~\ref{sec:momentmaps}) used in
  Fig.~\ref{fig:co65moments_memonly}.
    \label{fig:co65moments_compar}}
\end{figure*}

The moments maps extracted on the {\tt clean}, and {\tt uvmem}
datacubes are compared in Fig.~\ref{fig:co65moments_compar}. The
details of the centroid map do not depend on the
reconstruction. However, the intensity and velocity dispersion maps
inside the central beam (so within 0.2\arcsec~ of the star) result from
the superposition of several velocity components.

\section{Impact of a finite beam in estimates of intrinsic line-widths.} \label{sec:disp}

\begin{figure*}
\begin{center}
\includegraphics[width=0.8\textwidth,height=!]{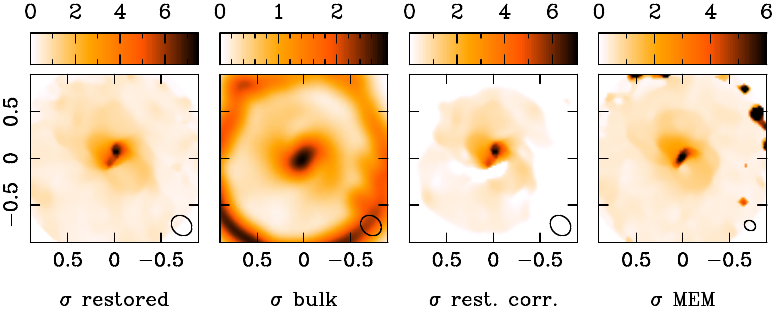}
\end{center}
\caption{Intrinsic line-width and structure within the central
  beam. All images are plotted in linear stretch. From left to right
  we show the restored velocity dispersion in Briggs weights,
  $\sigma_{\rm rest.}$, the bulk flow computed from the MEM centroid,
  $\sigma_{\rm bulk}$, the intrinsic line-width $\sigma_{\rm int.}$
  (labeled '$\sigma$ rest. corr.'), and the deconvolved line-width
  $\sigma_{\rm mem}$.} \label{fig:disp}
\end{figure*}

An alternative cross check on the quality of the deconvolved velocity
dispersion as an estimate of the intrinsic line-width, at least away
from the central beam, is to take statistics on the velocity centroid
$v_{\rm mem}$ extracted on the MEM datacubes. We degrade $v_{\rm mem}$
with the same beam as used in the Briggs restoration, $v^s_{\rm mem}
\equiv \langle v_{\rm mem} \rangle = v_{\rm mem} \otimes B$, and
calculate the velocity dispersion in the restored maps due to the bulk
flow with $\sigma_{\rm bulk} = \sqrt{ \langle v_{\rm mem}^2 \rangle -
  (v^s_{\rm mem})^2 }$, where $\langle v_{\rm mem}^2 \rangle \equiv
v_{\rm mem}^2 \otimes B$. We then subtract $\sigma_{\rm bulk}$ in
quadrature to estimate the intrinsic line-width: $\sigma_{\rm int.} =
\sqrt{\sigma_{\rm rest.} - \sigma_{\rm bulk}}$, where $\sigma_{\rm
  rest.}$ is the 1~$\sigma$ velocity dispersion extracted from the
restored datacubes. The corresponding dispersion at $\vec{P}$ is
$\sigma_{\rm int.}(\vec{P}) = 1.81~$km~s$^{-1}$. A few pixels have
slightly lower $\sigma_{\rm rest.}$ than $\sigma_{\rm bulk}$, which may
stem from the finite resolution of $v_{\rm mem}$, or to non-Gaussian
line profiles. As these pixels represent a small fraction, we set them
to zero. Fig.~\ref{fig:disp} compares the deconvolved velocity
dispersion, also shown in Fig.~\ref{fig:co65moments_memonly}, with the
dispersion of the MEM datacube after correction for bulk motion.

The $\sigma_{\rm rest.}$ map in Fig.~\ref{fig:disp} shows interesting
structure on scales much smaller than the clean beam. Two peaks in
velocity dispersion are very reminiscent of the orientation and
separation between HD~142527 and HD~142527B \citep[][]{Biller2012,
  Close2014ApJ...781L..30C}. The binary separation is 84~mas along
PA=127\degr, and the mass ratio is $<0.1$, where the inequality is due
to the correction for accretion luminosity on HD~142527B. Although it
is tempting to associate the secondary peak in CO(6-5) velocity
dispersion with HD~142527B, we see that it does not persist in the
deconvolved maps, while the main peak also shifts slightly in
position. These peaks must reflect features of the underlying velocity
field, so that the corresponding pixels mix blue and red shifted
emission in a coarse beam. 


{\it Facilities:} \facility{ALMA}.




\bibliography{/Users/simon/common/texinputs/merged.bib}

\end{document}